\documentclass[prb,aps, babel,twocolumn]{revtex4}
\usepackage{amsmath,amssymb,amstext,amscd,amsthm,amsopn,graphicx,indentfirst,mathrsfs}
\newcommand{\ka}{K_1\alpha^2}
\newcommand{\kb}{K_2\beta^2}
\newcommand{\Beff}{B_{eff}}
\newcommand{\Jpar}{J_{\parallel}}
\newcommand{\Jper}{J_{\perp}}
\newcommand{\f}{\phi}
\newcommand{\w}{\omega}

\renewcommand{\vec}[1]{{\bf #1}}
\graphicspath{{.}{./EPS/}}

\begin{document}

\title{Magnetothermal Transport in Spin-Ladder Systems}

\author{Ofer Shlagman}
%\email{ofer.shlagman@live.biu.ac.il}

\affiliation{Department of Physics, Bar Ilan University, Ramat-Gan 52900, Israel}
\author{Efrat Shimshoni}
%\email{shimshe@mail.biu.ac.il}

\affiliation{Department of Physics, Bar Ilan University, Ramat-Gan 52900, Israel}
\begin{abstract}
We study a theoretical model for the magnetothermal conductivity of a spin-$\frac{1}{2}$ ladder with low exchange coupling ($J\ll\Theta_D$) subject to a strong magnetic field $B$. Our theory for the thermal transport accounts for the contribution of spinons coupled to lattice phonon modes in the one-dimensional lattice. We employ a mapping of the ladder Hamiltonian onto an XXZ spin-chain in a weaker effective field $B_{eff}=B-B_{0}$, where $B_{0}=\frac{B_{c1}+B_{c2}}{2}$ corresponds to half-filling of the spinon band. This provides a low-energy theory for the spinon excitations and their coupling to the phonons. The coupling of acoustic longitudinal phonons to spinons gives rise to hybridization of spinons and phonons, and provides an enhanced $B$-dependant scattering of  phonons on  spinons. Using a memory matrix approach, we show that the interplay between several scattering mechanisms, namely: umklapp, disorder and phonon-spinon collisions, dominates the relaxation of heat current.
This yields magnetothermal effects that are qualitatively consistent with the thermal conductivity measurements in the spin-$\frac{1}{2}$ ladder compound ${\rm Br_4(C_5H_{12}N)_2}$ (BPCB).
\end{abstract}


%\keywords{Spin-chain, magnetothermal, thermal conductivity}
\maketitle

\small
\section{Introduction and Principal Results}

 Quasi one dimensional (1D) magnetic systems are present in a variety of new compounds with magnetic elements, and provide interesting manifestations of strongly correlated physics in electronic systems\cite{gia}. These systems are realized in crystals with a chain-like structure of the magnetic atoms, where intrachain exchange interactions are much stronger than interchain interactions. Their low dimensionality leads to the enhancement of quantum fluctuations, and the formation of exotic phases at low temperatures.

 In particular, spin-$\frac{1}{2}$ chain systems (most commonly realized in Cu-based compounds)\cite{zotos} are typically insulators in which the charge degree of freedom is frozen, and the  dynamics is restricted to the spin sector. The elementary excitations are  spin flips propagating along the chains direction. These can be described in terms of interacting Fermionic degrees of freedom, called spinons, which carry spin but no charge \cite{affleck}. These systems therefore provide one of the simplest realizations of Luttinger liquids (LL). This spinon LL is, in fact, the most abundant form of the so-called "spin-liquid" state, characterized by a magnetically disordered ground-state and power law spin-spin correlations\cite{balents-2010}. %with  the low-energy Hamiltonian
%\begin{equation}
%H_{LL}=\frac{u}{2\pi}\int dx[\frac{1}{K}(\partial_x\phi)^2+K(\pi\Pi)^2],
%\end{equation}
%where $u$ is a velocity and $K$ a dimensionless parameter.

% In terms of Bosonic operators, the interaction term of 1D Fermionic Hamiltonians can be simplified to give a quadratic Hamiltonian. The Bosonic operators are then related to Boson field operators, and

%Theoretical models describing these systems are solvable by powerful techniques specifically designed for one dimensional problems\cite{gia}, which is rarely possible in higher dimensional systems with strong interactions.
  The most elementary model for 1D spin systems is the XXZ Hamiltonian\cite{bethe}, describing a spin-$\frac{1}{2}$ chain with  nearest neighbor interactions,
 \begin{equation}
H_{XXZ}=\sum_iJ_{xy}(S_{i+1}^{x}S_{i}^{x}+S_{i+1}^{y}S_{i}^{y})+J_z\sum_iS_{i+1}^{z}S_{i}^{z}-B\sum_iS_i^z\;.
\label{eq:xxz}
\end{equation}
Here $J_{\alpha}>0$ corresponds to antiferromagnetic exchange interaction, and $B$ is an external magnetic field (note that here and throughout this paper we adopt units where $g\mu_{\rm B}=k_{\rm B}=\hbar=1$). The isotropic case $J_{xy}=J_z$ yields the 1D Heisenberg model.
  %called the  XXZ model. %when the interactions along the $x$ and $y$ directions are equal  and different than $z$ (i.e $J_x=J_y=J_{xy}\neq J_z$).
  %for isotropic interactions ($J_x=J_y=J_z=J$) it is called the  Heisenberg model.
On each site the spin operator is represented by $\vec{S}_i=\frac{\vec{\sigma}_i}{2}$ where $\sigma^{\alpha}$ are the pauli matrices. %The fact that different components of the spin operators do not commute illustrates the inherent quantum nature of the system. %As will be discussed below,
The spin chain can be  mapped into interacting spinless Fermions on a lattice\cite{gia,zotos}, where the magnetic field $B$ serves as a chemical potential. At zero field the Fermions are at half filling, and upon raising the magnetic field  they gradually polarize until saturation at $B=B_c$ which corresponds to a depletion of the spinon band.

 More complicated variants of the XXZ and Heisenberg model can describe quasi 1D systems with additional interactions such as zig-zag chains, spin-Peierls chains, and ladders\cite{gia,shelton,dagotto}. These systems support a richer phase diagram including, e.g, gapped dimer crystal phases. Upon tuning the magnetic field the system may undergo a phase transition from a gapped phase into a spin-liquid\cite{ladders}. In particular, in a ladder subject to a strong field $B_{c1}<B<B_{c2}$, a LL phase of gapless spinons is recovered\cite{orig}.

One of the prominent manifestations of a spin-liquid state is the contribution of gapless spinons to transport. Since there is no straightforward way to measure the spin current through an antiferromagnetic chain, investigation of the spinons properties can be done by measuring the thermal conductivity $\kappa$. Experimental evidence for a substantial enhancement of thermal conductivity  along the chains direction ($\kappa_{\parallel}$), has indeed been found in CuO based chain compounds \cite{spin-ex1,spin-ex2,spin-ex3}.  However, interpretation of the data is complicated by the dominant contribution of crystal phonons, and in particular their coupling to the spinons \cite{shimsh,rozhkov}. In principle, an obvious means of disentangling the spin degrees of freedom is the application of an external magnetic field $B$, which allows the tuning of system parameters in the spin sector only. The resulting {\em magnetothermal} effects -- namely, variations of $\kappa$ as a function of $B$ -- can serve as a valuable probe of the spin system. At low temperatures both spinons and phonons contribute to the heat transport. The total heat conductivity can be split into a pure phononic contribution, $\kappa_{ph}(T)$, and a magnetic part, $\kappa_{mag}(B,T)$. Then, we can extract the magnetothermal conductivity:
\begin{equation}
\Delta\kappa(B,T)=\kappa(B,T)-\kappa(0,T)=\kappa_{mag}(B,T)-\kappa_{mag}(0,T)\;.
\label{eq:dk}
\end{equation}

   Magnetothermal effects as mentioned above are practically inaccessible in the typical CuO compounds, where the large exchange coupling ($J$ of order $2000\;K$) dictates an enormous scale of the desired external field. In contrast, a field--tuned manipulation is easily accessible in organic based magnetic compounds, where $J$ is typically of order $10K$. An experiment in the organic spin-chain material ${\rm Cu(C_4H_4N_2)(NO_3)_2}$ \cite{solog1} measured the magnetothermal conductivity. It indicated a non-monotonic $B$-dependence of $\kappa_{\parallel}$, and in particular - a pronounced dip feature with a minimum at a field scale $B_{min}\sim T$. A subsequent theoretical study \cite{solog2} has shown that such feature arises due to the interplay between disorder and umklapp scattering of the spinons: the latter process is sensitive to the field-induced tuning of the spinon Fermi-level away from the middle of the band. It thus reflects the Fermionic character of the spinons.

  As opposed to  the spin-chain compounds mentioned above, in spin-{\em ladder} compounds, magnetothermal effects are expected to dominate at high $B$ where the spin-gap closes up. A recent experiment \cite{BPCB} measured the magnetothermal conductivity in the spin-ladder compound ${\rm Br_4(C_5H_{12}N)_2}$ (BPCB).  % A possible explanation may rely on 3D phonons that couple to the spinons via the rung exchange $J_{\perp}$, this coupling can result in hybridized spinon-phonon degrees of freedom that are not restricted to 1D.
An experimental study\cite{orig} of thermodynamic properties of  this compound confirmed that it is described  very well by the spin-ladder model  with $J_{\parallel}=3.6K$ (the exchange along the legs of the ladder) and $\;J_{\perp}\sim13K$ (the exchange along the rungs), and its appropriate LL representation in the gapless regime $B_{c1}<B<B_{c2}$.

Indeed, the experimental data of  Ref. [\onlinecite{BPCB}] indicate that upon raising the magnetic field, the  magnetothermal conductivity $\Delta\kappa(B)$  vanishes for fields smaller than $B_{c1}$. However, when the magnetic field is raised further and the spin-gap is closed,  there is a large {\it decrease} in the  magnetothermal conductivity. On top of this decrease there is a double dip feature  with a local maximum at $B_0=\frac{B_{c1}+B_{c2}}{2}$, corresponding to half-filling for the Fermionic excitations. We assert that this data can be qualitatively explained as follows: first, the spinons in this system are slower than  the phonons, therefore they act as  impurities  for the phonons \cite{Rasch_thesis}. This effect induces a decrease in the conductivity upon entering the spin-liquid regime ($B_{c1}<B<B_{c2}$). Second, around half-filling ($B=B_0$) there is  a positive spinonic contribution to transport  observed as a maximum at $B=B_{0}$ . The double dip feature resembles results obtained for spin-{\it chains}\cite{solog1,solog2} where the minimum in the magnetothermal conductivity corresponds to moving the chemical potential away from half filling, to a scale of order $T$.

Motivated by these observations, in the present paper we study a minimal model for the magnetothermal transport of a coupled spinon-phonon system in a single ladder. Our  theory  accounts for a crucial distinction between ladders and chains: the strong magnetic field required to enter the gapless spinons phase provokes an enhanced coupling  between spinons and phonons. This leads  to hybridization between the spinons and phonons excitations. In addition, scattering of the phonons by the slower spinons is magnified, generating a relatively strong negative contribution to $\Delta\kappa(B)$. Qualitatively, our calculated $\Delta\kappa(B,T)$ resembles the experimental data of  Ref. [\onlinecite{BPCB}].

The paper is organized as follows: in Sec. \ref{sec:model} we derive the low-energy model for the spin system in the presence of coupling to 1D phonons. In Sec. \ref{sec:transport} we study the effect of scattering processes on the thermal conductivity in the framework of the memory matrix approach for the calculation of the conductivity tensor, and obtain the leading magnetic field and temperature dependencies of the thermal conductivity $\kappa$. In Sec. \ref{sec:sum} we summarize and discuss the results. Finally, in appendices A through C we present details of the calculation of the various memory matrix elements.

\section{Low-energy Model for the coupled spin-phonon system}
\label{sec:model}

We wish to compute the thermal conductivity of a system which consists of antiferromagnetic spin-$\frac{1}{2}$ ladders interacting with the lattice phonons. To this end, we focus on a simplified model for such a system, which considers a single ladder - i.e., both spinons and phonons are one-dimensional. The parameters of the model are adjusted to mimic those of BPCB \cite{BPCB}, in particular assuming the limit $J_{\perp}>J_{\parallel}$ (strong rung coupling). In addition, we assume $J_{\parallel}\ll\Theta_D$. In this section we describe the low energy model of the system, and derive the eigenmodes which constitute the elementary excitations of the coupled spin-phonon system.
 \subsection{Bosonization of the spin ladder Hamiltonian}
 We begin by describing the spin system. The Hamiltonian of a spin-$\frac{1}{2}$ two leg ladder  in a magnetic field $B$ along the $z$-direction is

\begin{equation}
H_s=\sum_{i=1}^{N}\sum_{\nu=1}^2[J_{\parallel}\vec{S}_{i,\nu}\cdot\vec{S}_{i+1,\nu}-BS_{i,\nu}^z]+\sum_{i=1}^NJ_{\perp}\vec{S}_{i,1}\cdot\vec{S}_{i,2}\;,
\end{equation}
where $\nu=1,2$ denotes the leg index. For $J_{\perp}>J_{\parallel}$, it can be approximately mapped into an effective spin-$\frac{1}{2}$ chain in a weaker magnetic field\cite{milla} $B_{eff}=B-B_0$:
\begin{equation}\label{eq:eff chain}
H_s=\sum_{i=1}^N[J_{xy}^{eff}(\sigma_i^x\sigma_{i+1}^x+\sigma_i^y\sigma_{i+1}^y)+J_z^{eff}\sigma_i^z\sigma_{i+1}^z]-\sum_{i=1}^{N}B_{eff}\sigma_i^z\;,
\end{equation}
 %Where the pseudo-spin-$\frac{1}{2}$ operators $\sigma$ act on lower energy states $|S>_i$ and $|T_1>_i$ according to
%\begin{align}
%\sigma_{i}^{z}|S>_i&=-\frac{1}{2}|S>_i\;,&  \sigma_{i}^{z}|T_1>_i=\frac{1}{2}|T_1>_i\;,\nonumber\\
%\sigma_{i}^{+}|S>_i&=|T_1>_i,& \sigma_{i}^{+}|T_1>_i=0\\
%\sigma_{i}^{-}|S>_i&=0\;,&  \sigma_{i}^{-}|T_1>_i=|S>_i\;.\nonumber
%\end{align}
where the effective parameters are given by
\begin{align}\label{eq:eff par}
&J_{xy}^{eff}=\Jpar,\quad J_z^{eff}=\Jpar/2\;,\nonumber\\
&B_{eff}=B-B_0,\quad B_0\equiv \Jper+\Jpar/2\;.
\end{align}
The isospin operators $\sigma_i^{\alpha}$ describe the effective spin-$\frac{1}{2}$ dynamics characterizing the low energy sector, which at high $B$ is restricted to the singlet and lower triplet state on each rung. Hence, in distinction from the real-spin XXZ model [Eq. (\ref{eq:xxz})], $\langle\sigma_i^z\rangle=0$ corresponds to a time-reversal symmetry broken state. To derive the low-energy model for the dynamics of this system we first use the Jordan-Wigner transformation,
   \begin{equation}
\sigma_{i}^{+}\rightarrow c_{i}^\dag \exp(i\pi\sum_{j=-\infty}^{i-1}c_{j}^{\dag}c_j),\quad \sigma_i^z\rightarrow c_i^{\dag}c_i-1/2
\label{trans}
\end{equation}
 which  maps the spin problem onto a model of interacting spinless Fermions on a lattice:
\begin{widetext}
\begin{equation}
H_s=-t\sum_i(c_i^{\dag} c_{i+1}+h.c)+V\sum_i(c_{i}^{\dag}c_i-1/2)(c_{i+1}^{\dag} c_{i+1}-1/2)\;,
\label{eq:fermion}
\end{equation}
\end{widetext}
%-\mu\sum_i(c_{i}^{\dag}c_i-1/2)
where $t=J_{xy}^{eff}/2$ and $V=J_{z}^{eff}$. For $B_{eff}=0$ the Fermionic band is half-filled and the Fermi momentum is $k_F^{(0)}=\frac{\pi}{2a}$. Finite $B_{eff}$ corresponds to a chemical potential for the Fermions, which shifts the Fermi momentum into $k_F=k_F^{(0)}(1+M_{eff})$, with $M_{eff}$ an effective magnetization.

Near the middle of the band ($B_{eff}=0$), the Fermion operators can be expressed in terms of Bosonic ones related to the Fermion density fluctuations using the standard dictionary of abelian Bosonization (see, e.g., appendix D in Ref. [\onlinecite{gia}]). For the spin operators (in the continuum limit: $x=ia$) this yields
\begin{align}\label{eq:spin-boson}
&\sigma^+(x)=\frac{e^{-i\theta(x)}}{\sqrt{2\pi a}}[(-)^x+\cos(2\phi(x))]\; ,\nonumber\\
&\sigma^z(x)=-\frac{1}{\pi}\partial_x\phi(x)+\frac{(-)^x}{\pi a}\cos(2\phi(x))\;,
\end{align}
 where $\sigma^{\pm}(x)=\frac{1}{\sqrt{a}}\sigma_i^{\pm},\,\sigma^z(x)=\frac{1}{a}\sigma_i^z$ %for x=ia
 , and $a$ is the lattice constant. Substituting Eq. (\ref{eq:spin-boson}) into Eq. (\ref{eq:eff chain}) we can describe the low energy properties of the spin system in terms of the Boson Hamiltonian:
 \begin{align}\label{eq:boson hamiltonian}
 &H_s=H_s^0+H_u\nonumber\;,\nonumber\\
 &H_s^0=\frac{1}{2\pi}\int dx[g(\partial_x\phi(x))^2+v_F(\pi\Pi(x))^2]\;,\\
 &H_u=g_u\int dx\cos[4\phi(x)]\;,\nonumber
 \end{align}
 where
 \begin{equation}
 v_F=aJ_{\parallel},\quad g=v_F\left(1+\frac{2}{\pi}\right),\quad g_u=-\frac{J_{\parallel}}{4\pi^2a}\;,
\label{eq:boson parameters}
\end{equation}
and $\Pi(x)=\frac{1}{\pi}\partial_x\theta(x)$ is the canonical conjugate of $\f(x)$, obeying $[\Pi(x),\f(x')]=i\delta(x-x')$. $H_s^0$ is the standard LL Hamiltonian
 \begin{equation}
 H_{LL}=\frac{u}{2\pi}\int dx\left[\frac{1}{K}(\partial_x\f)^2+K(\pi\Pi)^2\right],
 \label{eq:LL ham}
 \end{equation}
where
 \begin{equation}
 u=v_F\left(1+\frac{2}{\pi}\right)^{1/2}\nonumber
 \end{equation}
 has the dimensions of velocity and
 \begin{equation}
 K=\left(1+\frac{2}{\pi}\right)^{-1/2}\nonumber
 \end{equation}
 is the dimensionless Luttinger parameter. Since $K>1/2$, the umklapp term $H_u$ is irrelevant (i.e. flows to zero under renormalization group (RG) for $T\rightarrow0$) and hence can be neglected in the description of the low-energy thermodynamic properties. However, as we shall see in the next section, it plays an essential role in the transport.

For $B\neq B_0$, the finite $B_{eff}$ introduces an additional term to $H_s^0$ due to the last term in Eq. (\ref{eq:eff chain}), which induces a finite effective magnetization. The most relevant correction is of the form
 \begin{equation}
\frac{1}{\pi}\int dx\Beff\partial_x\f\;,
\label{eq:Beff correction}
 \end{equation}
 which can be absorbed in the Gaussian part by a shift of the field $\f$, reflecting the shift of chemical potential for spinons. As implied by the exact Bethe ansatz solution, the LL form of $H_s^0$ is in any case maintained for arbitrarily large $\Beff$, but with renormalized parameters\cite{gia,haldane}. In particular $K(\Beff)$ approaches $1$ close to the edges of the band ($B\rightarrow B_{c1}$ or $B\rightarrow B_{c2}$).

 An additional correction to $H_s$ arises from weak disorder in the lattice, which can be accounted for by adding a random term $\delta B(x)$ to $B_{eff}$ in Eq. (\ref{eq:eff chain}). Such term may arise from defects leading to  random corrections to $J_{\perp}$, $J_{\parallel}$ via $B_0$ [Eq. (\ref{eq:eff par})]. This introduces a scattering term proportional to
 \begin{equation}
 \int dx\delta B(x)\cos[2\phi(x)]\;.
  \label{eq:dis correction}
  \end{equation}
  When we discuss the relaxation of the heat current, both the umklapp and disorder terms will become important, and will be considered as perturbations of $H_s^0$.\label{sec:bosonization}

  \subsection{Coupling to Lattice Phonon Modes}

Up to now we described only the spin system.  Next we will include the phonons in the model. In a single two-leg ladder of atoms, three modes of 1D phonons should be accounted for: two acoustic modes, longitudinal and transverse,  and an optical mode associated with fluctuations in the rung length.  %The effects of longitudinal  and transverse phonons on the system differ, the coupling of longitudinal phonons to the spin system leads to hybrid spinon-phonon degrees of freedom, while the coupling of transverse phonons leads to renormalization of the LL parameters $u,k$. For this reason, this subsection will focus on coupling to longitudinal phonons. The transverse phonons will be included in the model in the next subsection. We start by studying the coupling of longitudinal phonons to the spin system, %which give rise to hybrid spinon-phonon eigenmodes of the system. Then we turn to deal with transverse phonons which, as we show,their coupling to the spin-system renormalize the Luttinger liquid parameters $u,K$   %which add to the LL Hamiltonian of the spin system terms that couple between longitudinal
  The dominant coupling of phonons to spinons arises from the dynamical corrections to the exchange interaction, $J$, due to lattice vibrations. The spinons therefore couple to leading order only to the longitudinal acoustic mode (via fluctuations in $\Jpar$) and to the optical mode (via fluctuations in $\Jper$). %We first focus on the optical mode, which coupling to the spin sector merely normalizes the parameters.
% i took this term: -\sum_i\delta B_i\sigma_i^z. out of the previous equation right after the quadratic phonon part

We first consider the effect of coupling of optical transverse phonon modes to the spinons. We show that such coupling merely leads to normalization of the Luttinger parameters, $u$ and $K$ of Eq. (\ref{eq:LL ham}).

Let us define the transverse phonon field in the following way:
\begin{equation}
U_t(x)=U_{1}(x)-U_{2}(x),
\end{equation}
where $U_{1,2}$ are transverse displacements (along the rung direction) of atoms in different legs of a ladder normalized by the rung size $b$. Now, we substitute this definition of $U_t$ in the phonon-dependant exchange to obtain:
\begin{align}
&J_{\perp}(U_t)\approx J_{\perp}^{(0)}+\delta J_{\perp} U_t\;,\quad \delta J_{\perp}\equiv\left. b\frac{\partial J_{\perp}}{\partial r}\right|_{r=b}\;,\nonumber\\
&r=b(1+U_t).
\label{eq:exch1}
\end{align}
Inserting into Eqs. (\ref{eq:eff par}) and (\ref{eq:Beff correction}), we find that this adds to the Hamiltonian a term of the form
\begin{equation}
H_{sp}^t=-\delta J_{\perp}\int \frac{dx}{2\pi}U_t(x)\partial_x\f(x).
\end{equation}
It is useful to change into momentum representation with
\begin{widetext}
\begin{equation}
U_t(x)=\sqrt{\frac{a}{L}}\sum_k(b_ke^{ikx}+b_k^{\dagger}e^{-ikx}),\quad \f(x)=\sqrt{\frac{a}{L}}\sum_k(\f_ke^{ikx}+\f_{-k}e^{-ikx})
\end{equation}
\end{widetext}
(where $L$ is the length of the legs). Then, the quadratic part of the coupled spinon-phonon Hamiltonian is given by
\begin{equation}
H=H_s^0+H_p^t+H_{sp}^t,
 \end{equation}
 where $H_s^0$ describes the spinons in terms of a Luttinger Hamiltonian [Eq. (\ref{eq:LL ham})], %Eq. (\ref{eq:LL}),
 \begin{equation}
 H_p^t=\w_0\sum_kb_k^{\dag}b_k
  \end{equation}
  describes the optical transverse phonons, and
  \begin{equation}
  H_{sp}^t=-\delta J_{\perp}a\sum_kik(\f_kb_k^{\dag}-\f_{-k}b_k)
  \end{equation}
  is the spinon-phonon interaction. Using a coherent path integral representation, it is a straightforward exercise to integrate over the phonon degrees of freedom, yield an effective action for the spinons, $S_{eff}$, defined as
  \begin{equation}
  e^{-S_{eff}[\f]}=\int D\overline{b}Dbe^{-S[\f, \overline{b}, b]}.\nonumber
   \end{equation}
   In the limit $k,\;\w_n\rightarrow0$ this results in a Luttinger model with a modified coefficient of $(\partial_x\f)^2$:
\begin{equation}
\frac{u}{K}\rightarrow \frac{u}{K}+\frac{\delta J_{\perp}^2a}{\w_0}.
\end{equation}
Hence, the renormalized parameters become:
\begin{equation}
\tilde{K}=K\left[1+\frac{\delta J_{\perp}^2aK}{\w_0u}\right]^{-1/2},\quad\tilde{u}=u\left[1+\frac{\delta J_{\perp}^2aK}{\w_0u}\right]^{1/2}.
\end{equation}

Next we focus on the longitudinal phonons, which coupling to the spin sector has the most dramatic consequences. Assuming small displacements of atoms from their equilibrium positions, we can approximate the exchange interaction by:
\begin{equation}
J_{\parallel}(r)\approx J_{\parallel}^{(0)}+g_{\parallel}\partial_xU_l(x),\quad g_{\parallel}\equiv a^2\frac{\partial J_{\parallel}}{\partial r}|_{r=a},
\label{eq:exch2}
\end{equation}
where $r=a[1+U_l(x+a)-U_l(x)]$ is the distance between neighboring atoms on the same leg, and  the dimensionless field $U_l(x)$
describes the relative longitudinal displacements of atoms. When inserted into Eq. (\ref{eq:eff chain}), these corrections give rise to coupling between the spinons and phonons. The Hamiltonian describing longitudinal phonons traveling parallel to the chains is%\nopagebreak
%\begin{widetext}
\begin{equation}
H_p^l=\frac{v}{2\pi}\int dx[(\pi P_l(x))^2+(\partial_xU_l(x))^2],
\label{eq:phonon}
\end{equation}
%\end{widetext}
where $v\sim a\Theta_D$ (with $\Theta_D$ the Debye temperature) is the sound velocity, and  $P_l$ is the momentum conjugate to  $U_l$.% Here we assume for simplicity an identical velocity of both phonon modes

After inserting the phonon-dependant correction to the exchange interaction into Eqs. (\ref{eq:eff par}), (\ref{eq:boson hamiltonian}) and (\ref{eq:Beff correction}), and adding the phonon Hamiltonian [Eq. (\ref{eq:phonon})], the low-energy Hamiltonian of the coupled spin-phonon system can be written as:
\begin{align}
&H_0=\frac{1}{2\pi}\int dx\left\{g(\partial_x\tilde{\phi}(x))^2+v_F(\pi\Pi(x))^2+h_l\partial_xU_l(x)\partial_x\tilde{\phi}(x)\right.\nonumber\\
&\left.+v[(\partial_xU_l(x))^2+(\pi P_l(x))^2]+\frac{g_{\parallel}B_{eff}^{(0)}}{g}\partial_xU_l(x)\right\},
\label{eq:sp-ph hamiltonian}
\end{align}
 with
\begin{align}
&h_l\equiv-g_{\parallel}\left[1+\frac{2(1+2/\pi)B_{eff}^{(0)}a}{g}\right],\nonumber\\
&\partial_x\tilde{\phi}(x)\equiv\partial_x\phi(x)+\frac{\Beff^{(0)}}{g},\quad \Beff^{(0)}\equiv B-J_{\perp}^{(0)}-J_{\parallel}^{(0)}/2.
%\label{Beff_def}
\end{align}
In  Eq. (\ref{eq:sp-ph hamiltonian}) we neglected small terms (of order $\partial_xU_l(\partial_x\f)^2$ and higher). These terms are  irrelevant, and moreover correspond to forward scattering that cannot contribute to transport properties of the spinons to leading order. Note that, in contrast with spin-chains \cite{shimsh}, at half-filling ($B_{eff}^{(0)}=0$) the coupling to the phonons via the coupling constant $h_l$ is {\em linear} in the spinon field $\partial_x\phi$ and has to be included in the low-energy Hamiltonian. This reflects the breaking of time-reversal symmetry in the system, where spinons correspond to fluctuations around a partially polarized magnetic state. Below we show how these terms lead to new eigenmodes of mixed spinon-phonon degrees of freedom.\label{sec:coupling}

\subsection{Derivation of Hybrid  Eigenmodes}

The Hamiltonian $H_0$ in  Eq. (\ref{eq:sp-ph hamiltonian}) describes the low energy properties  of  the coupled spinon-phonon system. In order to find  the eigenmodes which constitute the elementary degrees of freedom  of the system, we proceed in diagonalizing  it by
a canonical transformation\cite{shimsh}:
\begin{align}\label{eq:bog1}
&\tilde{\phi}(x)=C\phi_1(x)-\lambda^2S\phi_2(x)\nonumber,\\
&U_l(x)=\frac{1}{\lambda^2}S\phi_1(x)+C\phi_2(x),
\end{align}
and similarly for the canonically conjugate momentum:
\begin{align}\label{eq:bog2}
&\Pi(x)=C\Pi_1(x)-\frac{1}{\lambda^2}S\Pi_2(x),\nonumber\\
&P_l(x)=\lambda^2S\Pi_1(x)+C\Pi_2(x),
\end{align}

where:
\begin{align}\label{eq:bog parameters}
&C\equiv\frac{1}{\sqrt{2}}\left[1-\frac{1}{\sqrt{A^2+1}}\right]^{1/2},\quad S\equiv\frac{1}{\sqrt{2}}\left[1+\frac{1}{\sqrt{A^2+1}}\right]^{1/2},\nonumber\\
&\lambda^2\equiv\sqrt{v_F/v},\quad A\equiv\frac{h_l\sqrt{v_F/v}}{gv_F/v-v}\approx-\frac{h_l}{v}\sqrt{\frac{v_F}{v}}\ll 1.
\end{align}
 Eqs. (\ref{eq:bog1}) and (\ref{eq:bog2}) are designed to preserve the canonical commutation relations $[\phi_{\nu}(x),\Pi_{\nu'}(x')]=i\delta_{\nu\nu'}\delta(x-x')$. The last approximation in Eq. (\ref{eq:bog parameters}) assumes $v_F\ll v$, which follows from $J_{\parallel}\ll \Theta_D$. The parameter $A$ defines the strength of  the coupling between the spinons and phonons. Note that it would be much stronger in a compound where $J\sim\Theta_D$, in which case the phonon and spinon velocities match, $v_F\sim v$.

  After this transformation, $H_0$ [Eq. (\ref{eq:sp-ph hamiltonian})] takes the form
\begin{align}
&H_0=\frac{1}{2\pi}\int dx\left\{\sum_{\nu=1,2}\left[\frac{v_{\nu}}{K_{\nu}}(\partial_x\phi_{\nu}(x))^2+v_{\nu}K_{\nu}(\pi\Pi_{\nu}(x))^2\right]\right.\nonumber\\
&\left.+\frac{h_l\Beff}{g}\left[\frac{S}{\lambda^2}\partial_x\phi_1(x)+C\partial_x\phi_2(x)\right]\right\},
\label{FnuPnu}
\end{align}
%\begin{align}\label{FnuPnu}
%&H_0=\frac{1}{2\pi}\int dx\{\frac{v_1}{K_1}(\partial_x\phi_1(x))^2\\
%&+v_1K_1(\pi\Pi_1(x))^2+\frac{v_2}{K_2}(\partial_x\phi_2(x))^2+v_2K_2(\pi\Pi_2(x))^2\nonumber\\
%&+v[(\partial_x\tilde{U}(x))^2+(\pi\tilde{P})^2(x)]+B^{eff}\frac{h}{g}\left[\frac{S}{\lambda^2}\partial_x\phi_1+C\partial_x\phi_2\right]\},
%\end{align}
with
\begin{widetext}
\begin{align}
&\frac{v_1}{K_1}=\frac{1}{2\sqrt{A^2+1}}\left[g(\sqrt{1+A^2}-1)+\frac{v^2}{v_F}(\sqrt{1+A^2}+1)+\sqrt{\frac{v}{v_F}}h_lA\right],\quad v_1K_1=v_F\nonumber\\
&\frac{v_2}{K_2}=\frac{1}{2\sqrt{A^2+1}}\left[\frac{gv_F}{v}(\sqrt{1+A^2}+1)+v(\sqrt{1+A^2}-1)-h_l\sqrt{\frac{v_F}{v}}A\right],\quad v_2K_2=v,\nonumber\\
\end{align}
\end{widetext}
which can be approximated for $v_F\ll v$ by
%\begin{widetext}
\begin{align}
&\frac{v_1}{K_1}\approx\frac{v^2}{v_F}\left(1-\frac{3A^2}{4}\right),\quad \frac{v_2}{K_2}\approx\frac{gv_F}{v}\left(1+\frac{3A^2v^2}{4gv_F}\right).
\end{align}
%\end{widetext}
In this form, $H_0$ is separable into two independent species of LLs. Using $A\ll 1$, the LL parameters are approximated by
\begin{align}
&v_1\approx v\left(1-\frac{3A^2}{4}\right)^{1/2}\; ,\quad v_2\approx v_F\sqrt{\left(1+\frac{2}{\pi}\right)\left(1+\frac{3A^2v^2}{4gv_F}\right)},\nonumber\\
&K_1\approx\frac{v_F}{v}\left(1+\frac{3A^2}{4}\right)^{1/2},\quad K_2\approx\frac{v}{v_F}\sqrt{\frac{1+\frac{3A^2v^2}{4gv_F}}{1+\frac{2}{\pi}}}.
\label{eq:luttinger par}
\end{align}
Finally, to get rid of the linear terms $\partial_x\phi_1(x)$ and $\partial_x\phi_2(x)$ in Eq. (\ref{FnuPnu}),
we define
\begin{align}
&\tilde{\phi}_1(x)=\phi_1(x)+\frac{B_{eff}K_1h_lS}{2v_1g\lambda^2}x\; ,\nonumber\\
&\tilde{\phi}_2(x)=\phi_2(x)-\frac{B_{eff}K_2h_lC}{2v_2g}x\; ,
\end{align}
and
\begin{align}
&\tilde{\Pi}_1(x)=\Pi_1(x),\quad\tilde{\Pi}_2(x)=\Pi_2(x)
\end{align}
which preserve the canonical commutation relations. The low energy Hamiltonian is now cast in the quadratic form of a LL:
\begin{equation}\label{eq:quadratic hamiltonian}
H_0=\sum_{\nu=1}^2v_{\nu}\int \frac{dx}{2\pi}\left[\frac{1}{K_{\nu}}(\partial_x\tilde{\phi}_{\nu}(x))^2+K_{\nu}(\pi\tilde{\Pi}_{\nu}(x))^2\right].
\end{equation}
%\begin{align}\label{eq:quadratic hamiltonian}
%&H_0=\frac{1}{2\pi}\int dx\left\{\frac{v_1}{K_1}(\partial_x\tilde{\phi}_1(x))^2+v_1K_1(\pi\tilde{\Pi}_1(x))^2\right.\nonumber\\
%&\left.+\frac{v_2}{K_2}(\partial_x\tilde{\phi}_2(x))^2+v_2K_2(\pi\tilde{\Pi}_2(x))^2\right\}.%+v[(\partial_x\tilde{U}(x))^2+(\pi\tilde{P}(x))^2]\}.
%\end{align}

The Hamiltonian (\ref{eq:quadratic hamiltonian}) is integrable (i.e. it has an infinite number of conservation laws), therefore the currents we are interested in (e.g heat current) are protected and cannot degrade. In order to get a finite conductivity we must add perturbations around $H_0$, e.g. the previously neglected umklapp term, which in terms of the shifted spinon field is given by
\begin{equation}
H_u=g_u\int dx\cos[4\tilde{\phi}(x)-\Delta kx],\;\Delta k\equiv\frac{4B_{eff}}{g}.
\label{eq:umklapp}
\end{equation}
$H_u$ describes processes where two spinons move from the right Fermi surface to the left (or vice versa), gathering momentum $ \Delta k=4k_F-G$ in which $G=\frac{2\pi}{a}$ is the reciprocal lattice momentum. Another important correction to $H_0$ is the backscattering term
\begin{align}
&H_{d}=\int dx\zeta(x)\cos[2\tilde{\phi}(x)],\;\zeta(x)\equiv\frac{\delta B(x)}{\pi a}\;,
%&H_{p-s}=g_{p-s}\int dx(\partial_xU)^2\partial_x\f\;,\nonumber\\
\label{eq:disorder}
\end{align}
  which describes scattering of spinons due to weak disorder caused by defects in the lattice. We assume uncorrelated random disorder where the sample average gives
 \begin{align}
 &\overline{\zeta(x)}=0,\nonumber\\
 &\overline{\zeta(x)\zeta(x')}=D\delta(x-x')\;.
 \label{eq:zetazeta}
 \end{align}
  Using Eq. (\ref{eq:bog1}), $H_u$ and $H_d$ can be expressed in terms of the hybrid spinon-phonon eigenmodes $\phi_1(x)$, $\phi_2(x)$:
 \begin{align}\label{eq:pert2}
&H_u=g_u\int dx\cos(2\alpha\phi_1(x)-2\beta\phi_2(x)-\Delta kx),\nonumber\\
&H_{d}=\int dx \zeta(x)\cos(\alpha\phi_1(x)-\beta\phi_2(x)),
\end{align}
with
\begin{align}\label{eq:ab}
&\alpha\equiv 2C,\quad\beta\equiv 2\lambda^2S\; .
%\Delta\tilde{k}\equiv\frac{4B_{eff}}{g_1}[1-\frac{Ah}{4\sqrt{1+A^2}}(\frac{\lambda^2}{F_2}-\frac{1}{F_1\lambda^2})]\\
\end{align}

   Finally, we note that higher orders in the expansions Eq. (\ref{eq:exch1}) and (\ref{eq:exch2}) yield an additional scattering term  between phonons and spinons, which turns out to have a significant effect on the transport carried by phonons. subsection \ref{sec:p-s} is devoted to a detailed study of the implication of this term on the thermal conductivity.  Together with $H_d,\; H_u$ [Eqs. (\ref{eq:umklapp}) and (\ref{eq:disorder})], this scattering process governs the degrading of currents leading  to a finite conductivity.

\section{Thermal Transport of the Spin-Ladder System}
\label{sec:transport}

The magnetothermal effects observed, e.g., in Ref. [\onlinecite{BPCB}] are a consequence of the interplay between different scattering mechanisms which result in the change of the thermal conductivity as a function of magnetic field $[\kappa(B)]$. Two primary effects are expected to dominate the $B$-dependance in the coupled phonon-spinon system: one arises from the positive contribution of spinons as heat carriers, and the other from the negative contribution of spinons acting as scatterers of the phonons. The former contribution is governed by the interplay of two scattering mechanisms, umklapp and disorder, and consequently depends on the deviation of the spinon chemical potential from a commensurate value\cite{solog2}. At the same time, the scattering due to phonon-spinon interaction is also dependant on the filling of the spinon band, which dictates the available phase-space for scattering. We note that due to the hybridization of phonons and spinons, these various effects are not entirely separable. In the present section we derive the magnetothermal conductivity of the spin-ladder model by using a memory matrix formalism, which allows an account of all the above mentioned scattering processes on equal footing.

\subsection{Approximate Conservation Laws and Memory Matrix Formalism}

To calculate the heat conductivity of the spin-ladder system, we use the memory matrix formalism \cite{forster} which has been successfully implemented in previous studies of thermal transport in spin-chains\cite{shimsh, solog2}. The memory-matrix approach is suited for systems where due to approximate conservation laws, the conductivity almost diverges \cite{AR}. The main step within this approach is the calculation of a matrix of relaxation rates for a given set of slow modes. The method allows to calculate transport coefficients within a hydrodynamic approximation, and provides a reliable lower bound to the conductivity \cite{Jung}. Moreover, it gives precise results as long as all the relevant slow modes are included in the calculation.

The heat transport properties of the spin--ladder system at low temperature are governed  by the approximate
conservation of a certain current, $J_c$ which has a finite overlap with the heat current. In particular, an exponentially slow decay of $J_c$ will lead to an exponentially large heat conductivity: the component of the heat current overlapping with $J_c$ is protected, and will decay exponentially slowly.

In the present case, the important step is to realize that in the absence of disorder, the linear combination
\begin{equation}\label{eq:Jc}
J_c=J_{\phi}-\frac{\Delta k}{4}J_s
\end{equation}
is conserved, meaning $[J_c,H_{0}+H_u]=0$, where $H_0$ is the LL Hamiltonian (\ref{eq:quadratic hamiltonian}), and $H_u$ the umklapp term [Eq. (\ref{eq:umklapp})]. Here,
 \begin{align}\label{eq:Js}
 &J_{\phi}=\int dx\Pi(x)\partial_x\phi(x),\nonumber\\
 &J_s=N_R-N_L=\int dx\Pi(x),
 \end{align}
 are the (normalized) heat current associated with the spinons and  the spin current, respectively, where $N_{R}$ and $N_{L}$ are the total number of right or left moving spinons. The overlap between the heat current and the conserved current is manifested by the appearance of $J_{\phi}$ in Eq. (\ref{eq:Jc}). The reason that $J_c$ is conserved by umklapp scattering is as follows: the umklapp term describes a process where a momentum $\Delta k$ is generated and therefore induces a change in $J_{\phi}$ proportional to $\Delta k$. In  the same process, the normalized spin current is changed by $-4$ as two right-moving spinons  are scattered into left-moving states. Since $J_c$ is conserved by the umklapp term $H_u$, the heat current can not be degraded by $H_u$ alone and additional scattering processes need to be accounted for.
 %A single umklapp process, therefore, cannot render the conductivity finite by itself \cite{andrei}.

The low-energy Hamiltonian $H_0$ conserves an infinite number of modes in addition to $J_c$. However, when perturbations are added, these modes decay faster than the conserved current $J_c$, since these modes do not commute with all the terms added.

We now  show how to calculate perturbatively the thermal conductivity when the relaxation of the heat current is dominated by the slow modes.  In our case the memory matrix is formulated in a space spanned by the slow modes $J_1$, $J_2,\;J_3$ and $J_s$ [Eq. (\ref{eq:Js})], where
\begin{align}\label{eq:currents}
&J_1=\int dx\Pi_1(x)\partial_x\phi_1(x),&J_2=\int dx\Pi_2(x)\partial_x\phi_2(x),\nonumber\\
&J_3=\int dx\Pi_3(x)\partial_x\phi_3(x)
\end{align}
which are all conserved by $H_0$. The fields $\phi_3,\Pi_3$ represent the transverse acoustic phonons which do not hybridize with spinons but are still scattered by spinons and therefore contribute to relaxation of the heat current. The heat current along the chains direction is $J_Q=v_1^2J_1+v_2^2J_2+v_3^2J_3$, where $v_1$ and $v_2$ are given in Eq. (\ref{eq:luttinger par}).

To set up the memory matrix formalism\cite{forster}, we first introduce a scalar product on the operators in the space spanned by the slow modes
\begin{equation}
(A(t)|B)=T\int_0^{1/T}d\lambda\langle A^{\dagger}(t)B(i\lambda)\rangle,
\end{equation}
where $\langle...\rangle$ denotes an expectation value at equilibrium, including average over disorder configurations.
The dynamic correlation function of the operators A and B is
\begin{align}
&\chi_{AB}(\omega)=\int_0^{\infty} dte^{i\omega t}(A(t)|B)\nonumber\\
&=\frac{iT}{\omega}\int_0^{\infty}dte^{i\omega t}<[A(t),B]>-\frac{(A|B)}{i\omega},
\end{align}
and the matrix of conductivities is given by

\begin{equation}\label{eq:sigma1}
\sigma_{pq}(\omega)=\frac{1}{TL}\chi_{J_pJ_q}(\omega)
\end{equation}
where $p,q$ are either of the slow modes. The heat conductivity is given by
\begin{equation}\label{eq:kappa1}
\kappa=\frac{1}{T}\sigma_{QQ}
\end{equation}
where $J_Q$ denotes the heat current. One can also write the matrix of {\em static} susceptibilities as
\begin{equation}\label{eq:susc}
\chi_{pq}=\frac{1}{TL}(J_p|J_q)\;.
\end{equation}
It can be shown \cite{forster} that the matrix of conductivities, $\hat{\sigma}$, can be expressed in terms of a memory matrix, $\hat{M}$
\begin{align}\label{eq:sigma2}
&\hat{\sigma}=\hat{\chi}(T) [\hat{M}(\omega,T)-i\omega\hat{\chi}(T)]^{-1}\hat{\chi}(T)\;.
\end{align}
The elements of the matrix $\hat{M}(\omega)$ in the d.c. limit ($\omega\rightarrow 0$) are
\begin{equation}\label{eq:memory}
M_{pq}=\lim_{\omega\to 0}\frac{C_{pq}(\omega)-C_{pq}(\omega=0)}{i\omega}\rightarrow -i\partial_{\omega}C_{pq}|_{\omega=0}
\end{equation}
where $p,\;q$  can be each of the slow modes of the theory, and $C_{pq}(\omega)$ is the Fourier transform of the retarded correlation function,
\begin{align}\label{eq:corr}
&C_{pq}(\omega)=\int_0^{\infty} dte^{i\omega t}\langle [F_p(t),F_q(0)]\rangle,
\end{align}
of the force operators:
%changed-check consistency
\begin{align}\label{eq:force}
&F_p\sim\dot{J}_p=i[H,J_p]=i[H^{pert},J_p].
\end{align}
Here $H^{pert}$ stands for perturbations to the low-energy Hamiltonian of the system $H_0$, that can relax the current $J_p$ [such as $H_u$, $H_d$, Eqs. (\ref{eq:umklapp}), (\ref{eq:disorder})].
 In the last equality we used $[J_p,H_{0}]=0$ for $p=s,1,2,3$, which justifies a perturbative expansion of $\hat{M}$: since $\dot{J}_p$ are already linear in perturbations around  $H_{0}$, the expectation values in Eq. (\ref{eq:susc}) and (\ref{eq:corr}) are computed with respect to $H_{0}$.

 From Eqs. (\ref{eq:kappa1}) and (\ref{eq:sigma2}) it follows that the d.c. thermal conductivity is given by
 \begin{equation}\label{eq:kappa2}
 \kappa=\frac{1}{T}\hat{\chi}\hat{M}^{-1}\hat{\chi}\;.
 \end{equation}
The static susceptibility matrix is given by
\begin{align}
\hat{\chi} = \left( {\begin{array}{*{20}c} \frac{2}{\pi v_F} & 0 & 0 & 0\\
0 & \frac{\pi T^2}{3v_1^3} & 0 & 0\\
0 & 0 & \frac{\pi T^2}{3v_2^3} & 0\\
0 & 0 & 0 & \frac{\pi T^2}{3v_3^3}
\end{array}} \right)
\end{align}
(where the matrix indices are $s,1,2,3$).
In our case the three leading perturbations contributing to the memory matrix, $\hat{M}$, are umklapp and disorder in the spin sector, and phonon scattering processes which include phonon-spinon interaction. Thus, the memory matrix is separated into three parts,
 \begin{equation}
 \hat{M}=\hat{M}^u+\hat{M}^d+\hat{M}^{p-s}\;.
 \end{equation}
Using the conservation law of the slow mode $J_c$ [Eq. (\ref{eq:Jc})] we find relations between the different umklapp matrix elements (see appendix A):
\begin{align}\label{eq:conservation}
&M_{s1}^u=\frac{\Delta k}{4}M^u_{ss}-M_{s2}^u,\nonumber\\
&M_{12}^u=\frac{\Delta k}{4}M_{s2}^u-M_{22}^u,\\
&M_{11}^u=\frac{\Delta k}{4}M_{s1}^u-M_{12}^u.\nonumber
\end{align}
 When $v_F\ll v$, and thus $\ka\ll \kb$ [see Eq. (\ref{eq:ab})], the matrix $\hat{M}^u$ greatly simplifies.
Using the relations (\ref{eq:conservation}) we find that the leading contribution to $\hat{M}^u$ depends only on one element $M_{ss}^u=M_s^u$, and we have
\begin{align}
\hat{M}^u \cong\left( {\begin{array}{*{20}c} M_s^u & 0 & \frac{\Delta k}{4}M_s^u & 0\\
 0 & 0 & 0 & 0\\
\frac{\Delta k}{4}M_s^u & 0 & (\frac{\Delta k}{4})^2M_s^u & 0\\
0 & 0 & 0 & 0
\end{array}} \right).
\end{align}
 The disorder contribution, $\hat{M}^d$, is a diagonal matrix with elements denoted by $M^d_{pp}=M^d_{p}$ ($p=s,1,2,3$). Finally, the dominant contribution from phonon-phonon and phonon-spinon scattering, $\hat{M}^p$, appears in the diagonal elements $M_1$, $M_3$ (a detailed calculation is provided in the next subsection).

Substituting these relations into Eq. (\ref{eq:kappa2}), we obtain an expression for the thermal conductivity
\begin{widetext}
\begin{align}\label{eq:kappa3}
 \kappa(B,T)\cong\frac{1}{T}\left[\frac{v_1^4\chi_{11}^2}{M_{1}}+\frac{16v_2^4\chi_{22}^2(M_s^{d}+M_s^u)}{\Delta k^2M_s^{d}M_s^u+16M_2^{d}(M_s^{d}+M_s^u)}+\frac{v_3^4\chi_{33}^2}{M_{3}}\right]\;.
 \end{align}
\end{widetext}
The $B$-dependance of this expression is encoded in the various matrix elements $M^u_s,\;M_s^d\;,M_1$ and $M_3$. We note that the scattering processes in the spin sector (described by $M_s^u,\;M_s^d$) dominate near half-filling of the spinon band, where their spectrum can be linearized and Bosonization is justified. Using Eq. (\ref{eq:pert2}) for the relevant terms in the Hamiltonian,
we derive $M_s^u$ (for a detailed calculation, see appendix A),
\begin{widetext}
\begin{align}\label{eq:Mus}
&M_{s}^u=g_{s}\left(\frac{T}{T_0}\right)^{2\kb+2\ka-3}B[\kb/2-i\delta,1-\kb]B[\kb/2+i\delta,1-\kb]\Re[\Psi(1-\kb/2-i\delta)-(\Psi(\kb/2+i\delta)]\nonumber\\
&g_{s}\equiv\frac{g_u^2}{2\pi^2}\sin(\pi \kb),\quad\delta\equiv\frac{v_2\Delta k}{4\pi T},\quad T_0\equiv\frac{v_2}{2\pi a},
\end{align}
\end{widetext}
where $B(x,y)=\frac{\Gamma(x)\Gamma(y)}{\Gamma(x+y)}$ is the Beta function, $\Psi(x)=\frac{\Gamma'(x)}{\Gamma(x)}$ is the digamma function and the parameters $K_1,\;K_2,\;\alpha$ and $\beta$ are defined in Eqs. (\ref{eq:luttinger par}) and (\ref{eq:ab}).
The dimensionless parameter $\delta$ determines the dominant field dependance of $\hat{M}^u$ via $\Delta k$ [see Eq. (\ref{eq:umklapp})]. For the  disorder part of $\hat{M}$ we find
\begin{widetext}
\begin{align}
\hat{M}^{d} =\left( {\begin{array}{*{20}c} D_{s} \left(\frac{T}{T_0}\right)^{\kb/2+\ka/2 -2} & 0 & 0 & 0\\
 0 & D_{1} \left(\frac{T}{T_0}\right)^{\kb/2+\ka/2} & 0 & 0\\
0 & 0 & D_{2}\left(\frac{T}{T_0}\right)^{\kb/2+\ka/2} & 0\\
0 & 0 & 0 & 0
\end{array}} \right)
\label{eq:Md}
\end{align}
\end{widetext}
%\begin{widetext}
% \[ \hat{M}^{d} =  \begin{pmatrix}
% D_{s} (\frac{T}{T_0})^{\kb/2+\ka/2 -2} & 0 & 0 & 0\\
% 0 & D_{1} (\frac{T}{T_0})^{\kb/2+\ka/2} & 0 & 0\\
%0 & 0 & D_{2}(\frac{T}{T_0})^{\kb/2+\ka/2} & 0\\
%0 & 0 & 0 & D_{3}\frac{T}{T_0} \end{pmatrix},\]
%\end{widetext}
with
 \begin{align}
 &D_{s}\cong\frac{4\pi Da^2}{ v_2^3},\;D_{1}\cong \frac{D\ka }{2v_1},\;D_{2}\cong \frac{ D\kb }{2v_2}
  \end{align}
where $D$ is defined in Eq. (\ref{eq:zetazeta}). The $B$-dependance of $\hat{M}^{d}$ is implicit in the parameters $K_1,\;K_2,\;\alpha,\;\beta$.

Incorporating Eqs. (\ref{eq:Mus}), (\ref{eq:Md}) in the second term of Eq. (\ref{eq:kappa3}), we find the primary magnetothermal effects originating from relaxation processes in the spin sector. To complete the derivation of $\kappa(B,T)$, one must account for the $B$-dependance of $M_1$, $M_3$ originating from phonon-spinon interaction. This part of the derivation requires special attention, and is discussed in subsection B below.

\subsection{Phonon-Spinon Scattering}
\label{sec:p-s}
%To this end we considered mainly the change in thermal conductivity as a result of changing the chemical potential of the spinons, thus we see a non monotonic behavior of the magnetothermal conductivity in the Spin-Liquid regime, but this effect does not complete the description of thermal transport in BPCB, on top of the double minimum feature there is a very large dip feature in the thermal conductivity which is explained as follows:%\subsection{Boltzmann Equation}

As already mentioned above, we focus on ladders with small exchange coupling obeying $J_{\parallel}\ll \Theta_D$, where the phonons typical velocity is  much larger than the spinons velocity. Therefore, the spinons act as impurities which scatter the phonons.   These scattering processes lead to relaxation of the phononic  heat current and therefore to a prominent dip in the thermal conductivity upon entering the partially filled spinon band, for $B_{c1}<B<B_{c2}$. These add a $B$-dependant contribution to the scattering rate of both longitudinal and transverse branches of acoustic phonons (represented by the Bosonic fields $U_l,\f_3$). Below we study the contribution of phonon-spinon scattering processes to the corresponding memory matrix elements.

In order to account for scattering processes of phonons on spinons, we expand the phonon-dependant exchange (\ref{eq:exch2}) to second order in $\partial_xU_l$. The most relevant phonon scattering term arising from this order of the expansion is of the form
\begin{align}
H_{p-s}=g_{p-s}\int\frac{dx}{2\pi}(\partial_xU_l)^2\partial_x\f.
\end{align}
A similar term resulting from the transverse displacements yields a coupling of spinons density to $(\partial_x\f_3)^2$.
We wish to consider magnetic-field dependance in a wide range, i.e. the entire spinon-band. For this purpose we model the spinons as free Fermions at a chemical potential dictated by $B$, a reasonable approximation, e.g., for the ladders in BPCB  where the Luttinger parameter $K$ is not far from 1 for arbitrary $B$ [\onlinecite{orig}]. Using $\partial_x\f(x)=-\pi\rho(x)$ [\onlinecite{gia}], and turning to Fourier space, $H_{p-s}$ acquires the form:
\begin{align}
&H_{p-s}=\sum_{kk'pp'G_n}V_{pp'}c_{k'}^{\dag}c_kb_{p'}^{\dag}b_p\delta(k-k'+p-p'-G_n),\nonumber\\
&V_{pp'}=\frac{g_{p-s}pp'}{\sqrt{\w_p\w_{p'}}},
\end{align}
where $c_k$ is a Fermionic (spinon) annihilation operator, and $b_{p}$ is a Bosonic (phonon) annihilation operator. $H_{p-s}$ describes elastic scattering where a phonon and a spinon with momenta $p$ and $k$ respectively, scatter into $p'$ and $k'$ respectively.

\begin{figure}[h]
\begin{center}
\includegraphics[width=1.0\linewidth]{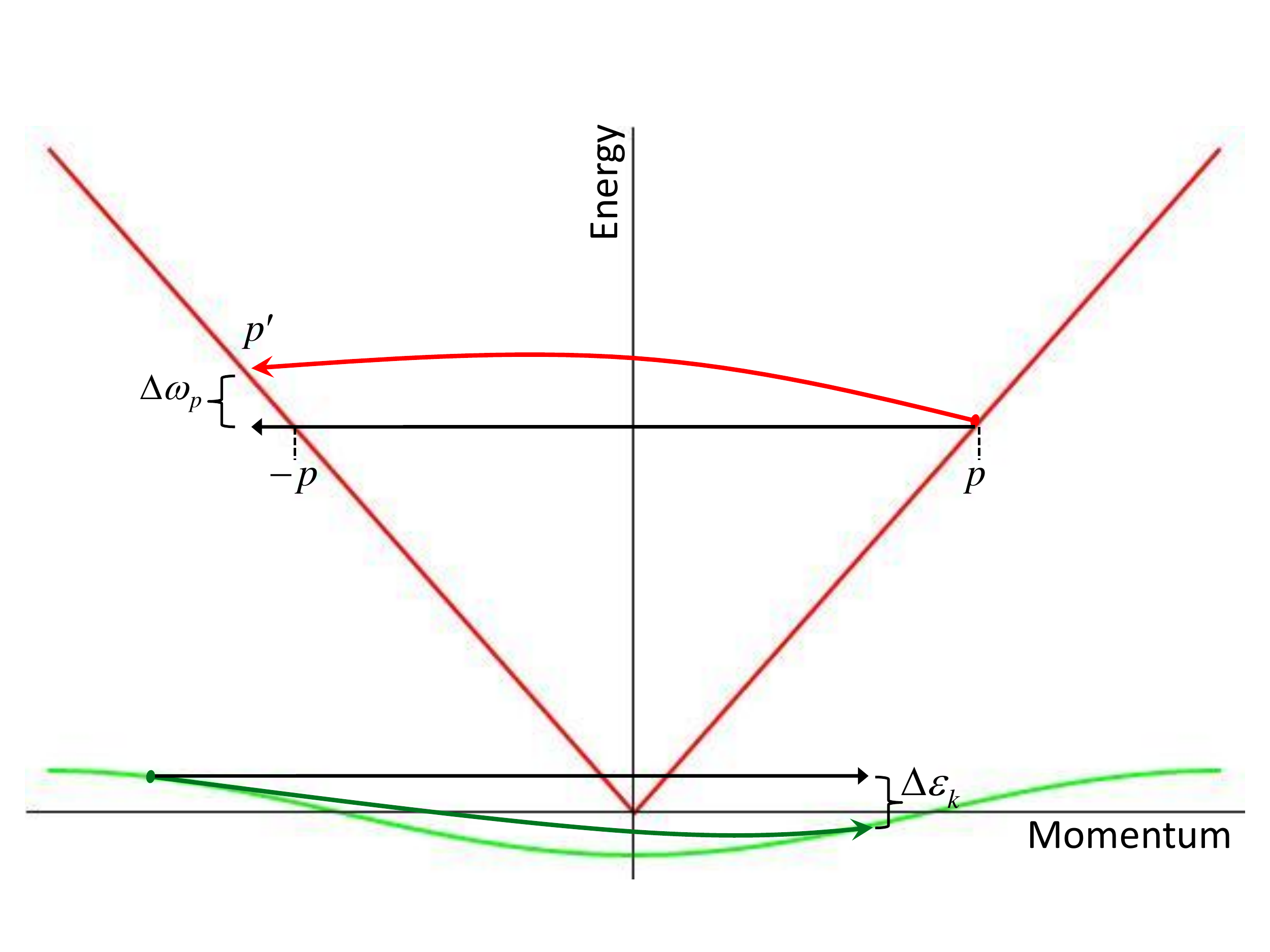}
\caption{(color online) Schematic Plot of the phonons (red) and the spinons (green) energy dispersions, the allowed phonon backscattering (red curved arrow) and the corresponding spinon scattering (green curved arrow). The black arrows represent the momentum transfer $q_0=p'-p=k-k'$.}
\label{fig:dispersions}
\end{center}
\end{figure}

\begin{figure}[h]
\begin{center}
\includegraphics[width=1.0\linewidth]{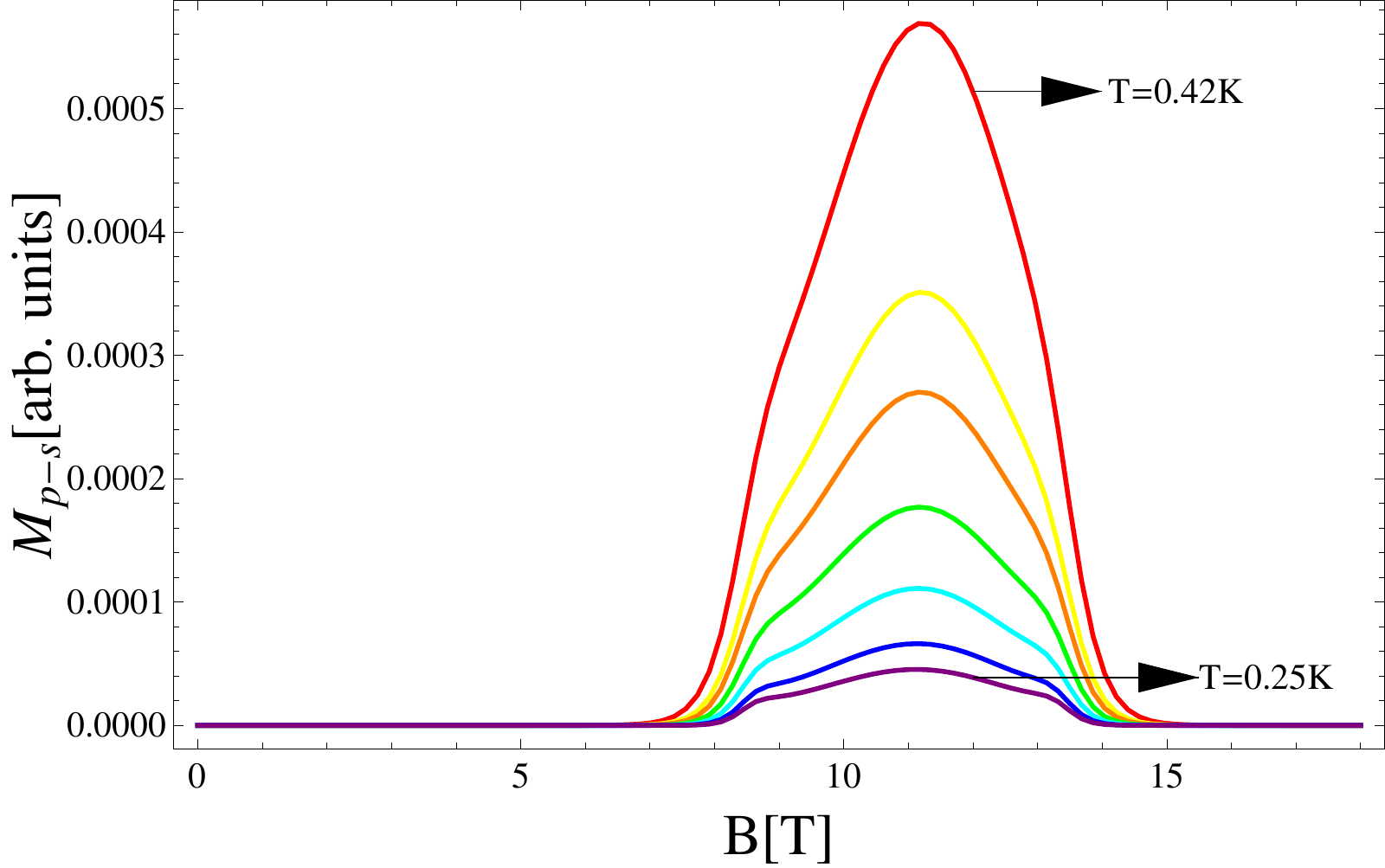}
\caption{(color online) Isotherms of $M_{p-s}$ as a function of magnetic field for various temperatures $T=0.25K,\,0.27K,\,0.30K,\,0.33K,\,0.36K,\,0.38K,\,\,0.42K$. The parameters used for this plot: $J_{\parallel}=3.6K,\,J_{\perp}=14.4K,\,g_{p-s}=0.04$, $\frac{v}{a}=18K$.}
\label{fig:Mps iso}
\end{center}
\end{figure}

\begin{figure}[h]
\begin{center}
\includegraphics[width=1.0\linewidth]{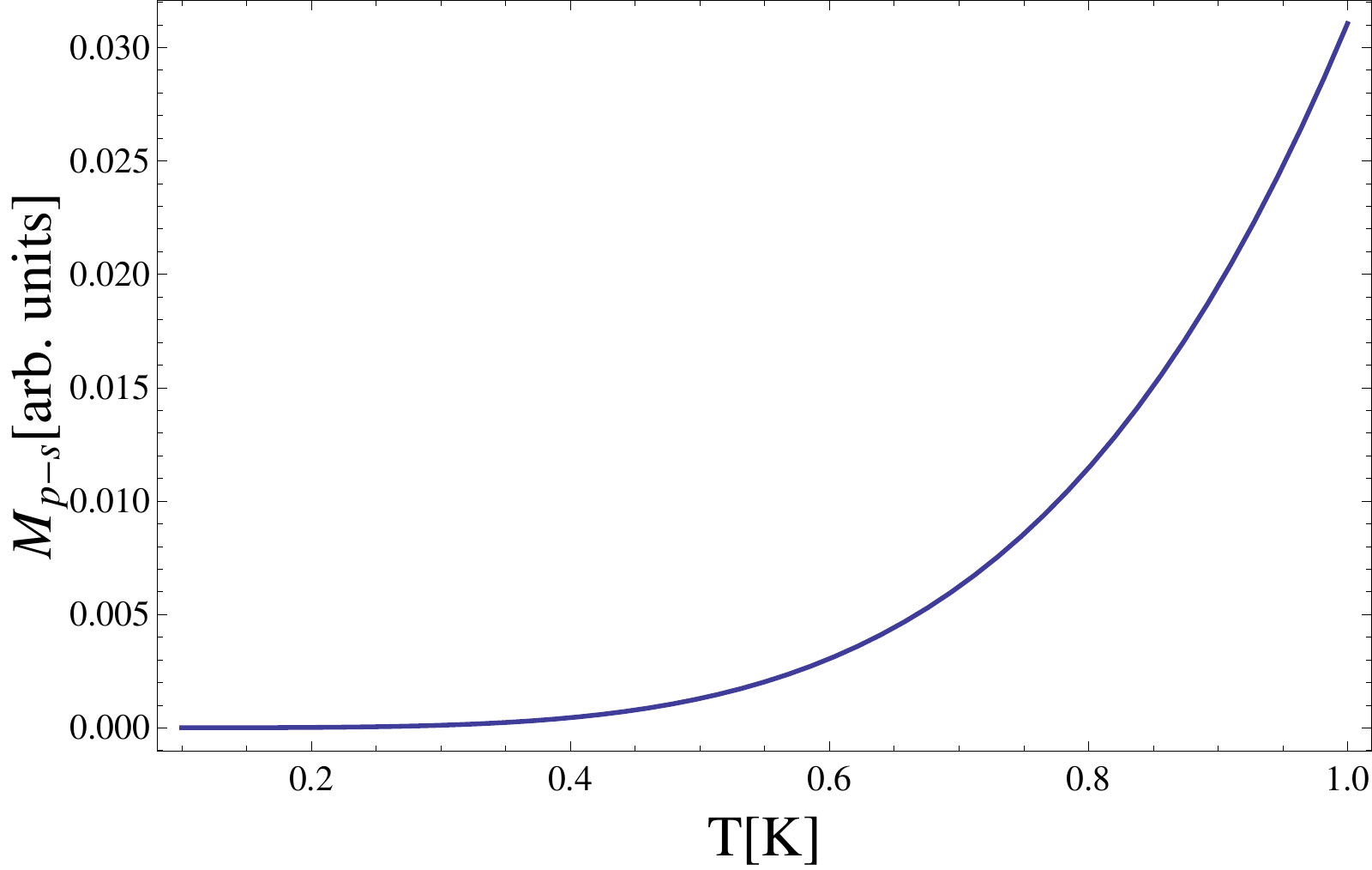}
\caption{(color online) $M_{p-s}$ at half-filling  ($B_{eff}=0$) as a function of temperature. The parameters used for this plot: $J_{\parallel}=3.6K,\,J_{\perp}=14.4K,\,g_{p-s}=0.04,\,\frac{v}{a}=18K$.}
\label{fig:TMps}
\end{center}
\end{figure}

We need to calculate the effect of this term on the phononic heat current
\begin{equation}
J_{ph}^E=\sum_pv_p\w_pb_p^{\dag}b_p,\quad v_p=\frac{\partial\w_p}{\partial p}\approx v\,\text{sign}(p),
\end{equation}
where we have assumed a linear dispersion $\omega_p\approx v|p|$.
 %the memory matrix element for phonon-spinon scattering is
%\begin{equation}
%M_{p-s}=\lim_{\omega\to 0}\frac{C_{p-s}(\omega)-C_{p-s}(\omega=0)}{i\omega}\rightarrow -i\partial_{\w}C_{p-s}|_{\w=0}\;,
%\label{eq:Mps}
%\end{equation}
The memory matrix element [Eq. (\ref{eq:memory})] is therefore calculated with the correlation function (\ref{eq:corr})
%\begin{equation}
%C_{p-s}=\langle F_{p-s}(t);F_{p-s}(0)\rangle,
%\end{equation}
of the force operator
\begin{equation}
F_{p-s}\sim[J_{ph}^E,H_{p-s}].
\end{equation}
Using Wick's theorem  we obtain an expression for this correlation function %appearing in (\ref{eq:memory})
(see appendix C for details)
\begin{align}
&C_{p-s}=\sum_{kpq} W_{pq}\delta(\Delta\w)n_{p+q}f_{k-q}(1+n_p)(1-f_k),\nonumber\\
&W_{pq}= -2v^2g_{p-s}^2|p(p+q)|q^2,\quad q\equiv p'-p,\nonumber\\
&\Delta\w\equiv\epsilon_k-\epsilon_{k-q}+\w_p-\w_{p+q},
\end{align}
where $f_k=(e^{(\epsilon_k-\mu)/T}+1)^{-1},\; n_p=(e^{\w_p/T}-1)^{-1}$, are Fermi and Bose  distributions respectively.
The memory matrix element is the derivative of $C_{p-s}$ with respect to $\w$. Integrating by parts and using the energies delta function we obtain:
\begin{widetext}
\begin{align}
&M_{p-s}=\int dkdp\frac{\partial}{\partial q}\left[\frac{W_{pq}n_{p+q}f_{k-q}(1+n_p)(1-f_k)}{\frac{\partial\Delta\w}{\partial q}}\right]_{q=q_0}\frac{1}{\frac{\partial\Delta\w}{\partial q}|_{q=q_0}},
\label{eq:Mps1}
\end{align}
\end{widetext}
%_{-\frac{\pi}{a}}^{\frac{\pi}{a}}
where $q_0$ is the momentum  transfer that obeys the energy and momentum conservation dictated by the delta functions. Since the phonon dispersion $\w_p=v|p|$ is much steeper than the spinon dispersion $\epsilon_k=-J_{\parallel}^{(0)}\cos(ka)$ , energy and momentum conservation can only be satisfied by phonon backscattering where $p\rightarrow p+q=-p+\delta p,\;|\delta p|<<|p|$. This is because a small change in phonon momentum will lead to a  large energy change, while the spinon energy transfer is small for small momentum transfer (see Fig. \ref{fig:dispersions}).
The integrals in  Eq. (\ref{eq:Mps1}) were solved numerically after approximating
\begin{equation}
\delta p\cong \frac{J_{\parallel}^{(0)}}{v}[\cos(k+2p)-\cos(k)]\; , \nonumber
\end{equation}
to get the temperature and field dependencies of $M_{p-s}$ (see Figs. \ref{fig:Mps iso}, \ref{fig:TMps}). The memory matrix is closely related to the relaxation time of the scattering process $M_{p-s}\sim\tau_{p-s}^{-1}$. Indeed as indicated by Fig. \ref{fig:Mps iso}, phonon scattering occurs practically only for $B_{c2}<B<B_{c1}$, in the spin-liquid phase where the spinons are gapless, and it is maximal for $B=B_{0}$ (half filling of the spinon band).
The temperature dependance of $M_{p-s}$ (Fig. \ref{fig:TMps}) for $B_{eff}=0$ gives a good fit to a power law $M_{p-s}\sim T^\sigma$, with $\sigma\cong 4.5$.

\begin{figure}[h]
%\begin{minipage}[b]{0.5\linewidth}
\centering
\includegraphics[width=1.0\linewidth]{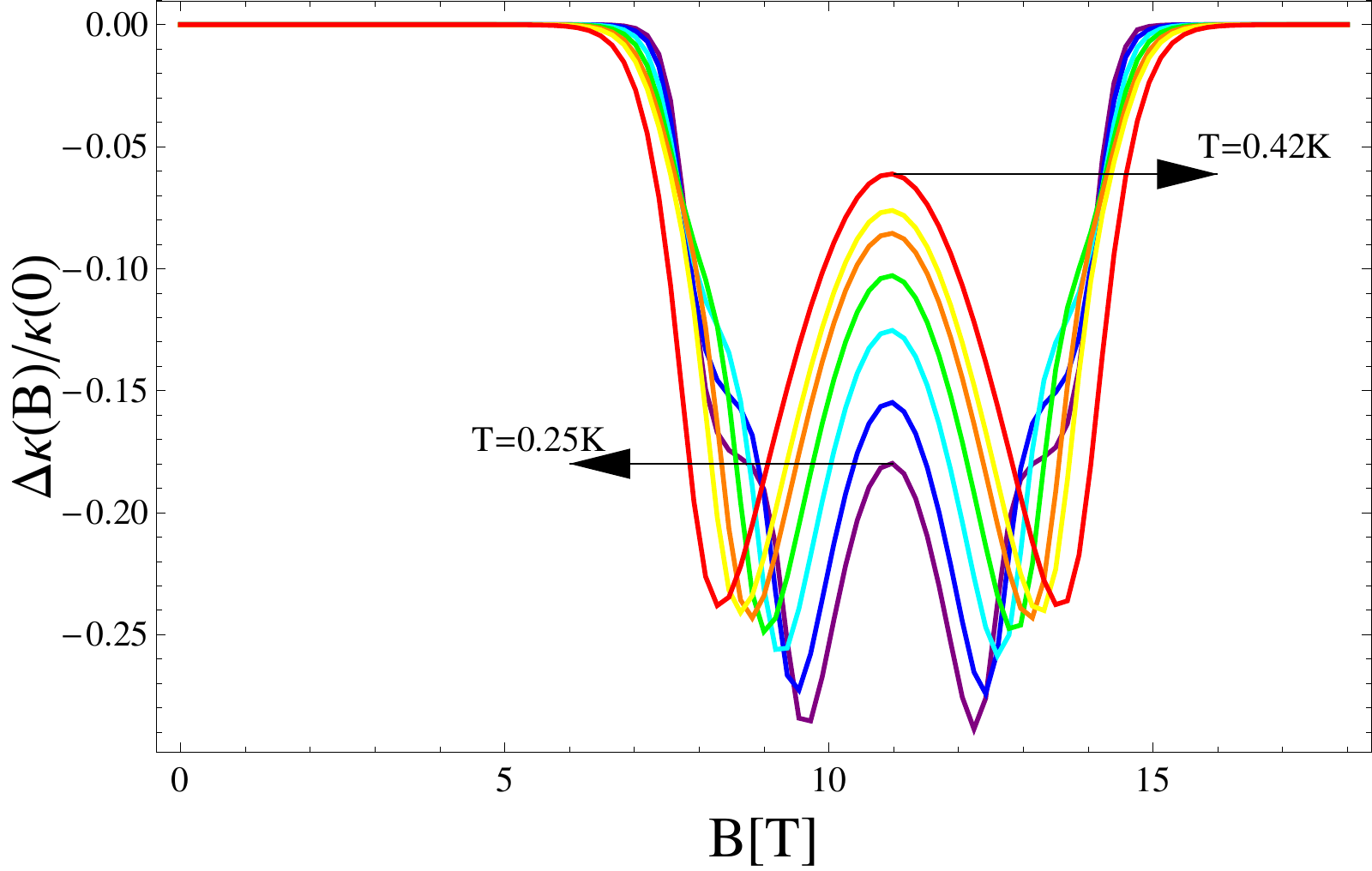}
\caption{(color online) Isotherms of the normalized magnetothermal conductivity as a function of magnetic field for various temperatures $T=0.25K,\,0.27K,\,0.30K,\,0.33K,\,0.36K,\,0.38K,\,\,0.42K$. The parameters used for this plot: $J_{\parallel}=3.6K,\,J_{\perp}=14.4K,\,g_{p-s}=0.04,\,D=1.6\times10^{-8},\,\,\frac{v}{a}=18K,\, D_p=7.4\times10^{-8},\,D_3=4.6\times10^{-6},\,\gamma=4$.}
\label{fig:boltzmann1}
%\end{minipage}
\hspace{0.5cm}
%\begin{minipage}[b]{0.5\linewidth}
%\centering
%\includegraphics[height=150pt,width=150pt]{kappa_isoterms_article23102011_with_boltzman_analytic}
%\caption{Theoretical result of the normalized magnetothermal conductivity as a function of magnetic field, with Boltzmann equation solved analytically.}
%\label{fig:boltzmann2}
%\end{minipage}
\end{figure}

We next recall that under the approximation $v_F\ll v$, one obtains $\alpha\ll\beta$ [see Eqs. (\ref{eq:bog parameters}), (\ref{eq:ab})], which implies that one hybrid mode $\f_1$ is phonon-like, while $\f_2$ is spinon-like. Therefore, the scattering of longitudinal phonons is included to a good approximation only in the $M_1$ element of  the memory matrix. This is added to the disorder term already retrieved earlier, and a $B$-independent contribution which assumes a power-law dependance on $T$. We thus obtain an expression of the form
\begin{equation}\label{eq: M1}
M_1=D_1\left(\frac{T}{T_0}\right)^{\kb/2+\ka/2}+D_pT^{\gamma}+M_{p-s}\; .
\end{equation}
A similar expression, excluding the first term, holds for $M_3$ which describes the scattering of transverse acoustic phonons.
Substituting in (\ref{eq:kappa3}), we obtain the final expression for $\kappa(B,T)$ and consequently for $\Delta\kappa(B,T)$ [Eq. (\ref{eq:dk})]. The resulting $B$ and $T$ dependance of $\Delta\kappa(B)/\kappa(0)$ are plotted as a function of magnetic field for different temperatures (Fig. \ref{fig:boltzmann1}).
We note that this result, although based on a highly simplified minimal model which captures the main physics of the system, qualitatively reproduces the prominent features of the experimental data of Ref. \onlinecite{BPCB}.

\section{Summary and Discussion}
\label{sec:sum}
In this work we studied the thermal conductivity of weakly disordered spin ladders subject to a magnetic field and coupled to phonons. %, with an intention to explain the magnetothermal effects seen in BPCB \cite{BPCB}.
We found that due to coupling between the  phonons and the spins, the elementary degrees of freedom are hybrid spinon-phonon modes and strong scattering of phonons on spinons is induced. Our study of the phonon-spinon scattering found that due to energy and momentum conservation only certain {\em backscattering} processes are allowed. The phonon-spinon scattering along with umklapp and disorder scattering lead to a prominent dip in the thermal conductivity.
%($\sim25\%$ decrease).
We examined the mechanisms responsible for the relaxation of the heat current, and showed that an interplay between umklapp, weak disorder and phonon-spinon scatterings underlies the transport properties at low temperatures. For this system it leads to  minima in the thermal conductivity isotherms when the effective field is of the order of the temperature $|B_{eff}|\sim T$, while a local maximum appears for zero effective field, when $B=B_{0}$. %Since we mapped the ladder into an effective chain,
In the vicinity of $B_{0}$ there is a single dimensionless parameter $\delta$ which determines the leading field and temperature dependencies of the thermal conductivity. $\delta$ depends on the field via the momentum $\Delta k$ [Eq. (\ref{eq:umklapp})]: by substituting $\Delta k$  into $\delta$ [Eq. (\ref{eq:Mus})] we obtain an approximate (for $v_2\sim v_F$) expression for $\delta$:
\begin{equation}
\delta\equiv\frac{v_2\Delta k}{4\pi T}=\frac{v_2B_{eff}}{gT}\sim \frac{B_{eff}}{T}\; .
\end{equation}
 These features can be compared with the effects seen in chains \cite{solog2} (by interchanging $B_{eff}$ and $B$) where the single minimum is at a field $B\sim T$ and the maximum at $B=0$. Our results for the thermal conductivity isotherms (Fig. \ref{fig:boltzmann1}) display similar field and temperature dependance to those measured in the experiment\cite{BPCB}. %The interesting features can be explained as follows, for small fields $B<B_{c1}$  there is a gap to excitations, hence the magnetothermal conductivity vanishes. At $B=B_{c1}$ the gap closes and magnetothermal effects are observed.   In the case of zero effective field when $B=\frac{B_{c1}+B_{c2}}{2}$, $\Delta k=0$, as a result umklapp scattering plays no role and the heat conductivity is maximal corresponding to the half filled Fermion band.

It should be emphasized that our model relies on some simplifying assumptions, and most importantly focuses
on a purely 1D system corresponding to a single ladder. To account for the perpendicular magnetothermal
effects measured in the experiment \cite{BPCB}, our model should be extended to include phonons traveling perpendicular to the chains direction. Taking into account the coupling of such phonons with the spin ladders could result in hybrid spinon-phonon degrees of freedom with higher-dimensional dynamics. Hence, due to this hybridization we expect to obtain a higher dimensional spin-liquid-like state with strong anisotropies which will account for the perpendicular magnetothermal transport.

An additional limitation on the applicability of our theory to a realistic system is that we have assumed a naive model for the disorder, and in particular treat it perturbatively. This approximation breaks down at sufficiently low $T$: the disorder being a relevant perturbation eventually leads to localization, and an effective breaking of the ladders to weakly coupled segments of finite length \cite{solog1,BPCB}.

Finally, it should be noted that we have implemented an approximate mapping of a ladder onto a chain \cite{milla} which amounts to the truncation of high energy triplet states, and is formally justified for $J_\perp\gg J_\parallel$. Coupling to the high energy sector is likely to induce asymmetry between positive and negative deviations of $B$ from $B_0$, as indeed observed in the experiment \cite{BPCB}.

As a concluding remark, in this work we focused on the limit $v_F\ll v$, compatible with the parameters of BPCB. However, in other quasi 1D spin compounds, where $J\sim\Theta_D $ (e.g., ${\rm NaV_2O_5}$ [\onlinecite{NaVO}] and NO[Cu(NO$_3$)$_3$] [\onlinecite{CF}]), the spinons and phonons velocities are comparable in size $v_F\sim v$. Hence a strong hybridization between the two degrees of freedom is expected in such compounds. Our theoretical approach can be extended to account for this phenomenon as well; we expect to investigate it further in future work.

\acknowledgements

We gratefully acknowledge illuminating discussions with N. Andrei, T. Giamarchi, J. A. Mydosh and D. Podolsky, and particularly with D. Rasch, A. Rosch and A. V. Sologubenko. E. S. is grateful to the hospitality of the Aspen Center for Physics (NSF
grant 1066293). This
work was supported by the Israel Science Foundation (ISF) grant
599/10.

\appendix

\section{Umklapp Memory Matrix}

\label{sec:app1}
%In the first section of the appendix we evaluate only the umklapp elements of the memory matrix, and the second section will be devoted to the disorder part, having said that, we omit the index $u/d$ throughout this section.
Before proceeding into the calculations  of correlation functions, we show that due to the conservation law (\ref{eq:Jc}), simple relations between the umklapp matrix elements can be found. Substituting Eqs. (\ref{eq:bog1}), (\ref{eq:bog2}) into (\ref{eq:currents}) and using Eq. (\ref{eq:force}), we get

\begin{equation}
J_1+J_2=J_{\phi}+J_{U}
\end{equation}
where $J_U$ is the longitudinal phonons current $J_U=\int dxP_l\partial_xU_l$.
In addition we have
\begin{equation}
F_U=[J_U,H_u]=0\,,\quad
\Rightarrow F_1+F_2=F_{\phi}\; .
\end{equation}
Substituting this into the conservation law (\ref{eq:Jc}), we find
\begin{equation}
F_1+F_2=\frac{\Delta k}{4}F_s.
\end{equation}
Then it is easy to see, from $M_{pq}\sim\langle F_p;F_q\rangle$, the following relations:
\begin{align}\label{eq:cons}
&M_{s1}^u=\frac{\Delta k}{4}M^u_{ss}-M_{s2}^u,\nonumber\\
&M_{12}^u=\frac{\Delta k}{4}M_{s2}^u-M_{22}^u,\\
&M_{11}^u=\frac{\Delta k}{4}M_{s1}^u-M_{12}^u.\nonumber
\end{align}

According to Eq. (\ref{eq:memory}) and (\ref{eq:corr}), we need to calculate the Fourier transform of retarded correlation functions of the form
\begin{align}
&C_{pq}^u(x,t)=\langle f_p^u(x,t);f_q^u(0,0)\rangle_0,
\end{align}
with the force density operators $f_p^u(x,t)$ defined so that
\begin{align}
F_p^u= i[J_p,H_u]\equiv \int\, dxf_p^u(x),
\end{align}
in which $H_u$ is the umklapp term defined in Eq. (\ref{eq:umklapp}). The expectation value $\langle ... \rangle_0$ is evaluated with respect to $H_0$ (\ref{eq:quadratic hamiltonian}).
The first umklapp term  to  calculate is $M^u_{ss}$; from commutator identities we find
\begin{align}
&f_s^u(x)=i[\Pi(x),H_{u}]=\\
&ig_u\int dx^\prime[\Pi(x),\cos[4\phi(x^\prime)-\Delta kx^\prime]]=-4\pi g_u \sin[4\phi(x)-\Delta kx]\;.\nonumber
\end{align}

 To calculate correlation functions between trigonometric functions, we use the result (appendix C in [\onlinecite{gia}]):
\begin{equation}\label{eq:expcorr}
\langle\prod_j e^{iA_j\phi(r_j)}\rangle=e^{-\frac{1}{2}K\sum_{i<j}A_iA_jK\mathcal{F}_1(r_i-r_j)},
\end{equation}
 with $A_i$ some constants, $K$ the LL parameter, and $\mathcal{F}_1=\frac{1}{2}\ln\left[\left(\frac{\beta u}{\pi a}\right)^2\left(\sinh^2\left(\frac{\pi x}{\beta u}\right)+\sin^2\left(\frac{\pi\tau}{\beta}\right)\right)\right]$, this correlation function has the property that for  $\sum_iA_i\neq0$, it  equals zero.  Since $H_0$ is separable in terms of the eigenmodes $\phi_i$ ($i=1,2$), the correlation function $C^u_{ss}$ can be written as a product of two correlation functions
 \begin{widetext}
 \begin{align}
 &C_{ss}^u(x,t)= C_1(x,t)C_2(x,t)\equiv\langle e^{i[2\alpha\phi_1(x,t)-2\alpha\phi_1(0,0)]}\rangle\langle e^{i[2\beta\phi_2(x,t)-2\beta\phi_2(0,0)]}\rangle,
 \end{align}
 \end{widetext}
where the correlation function of each species of the eigenmodes  is calculated independently with respect to the corresponding LL Hamiltonian. This yields:
\begin{widetext}
\begin{align}
&C_{ss}^u(x,t)\cong \sin(\pi K_2\beta^2)\left(\frac{T}{T_0}\right)^{2K_2\beta^2+2\ka}(\sinh[\pi T(x/v_1-t+i\epsilon)]\sinh[\pi T(x/v_1+t-i\epsilon)])^{-\ka}\times\nonumber\\
&(\sinh[\pi T(x/v_2-t+i\epsilon)]\sinh[\pi T(x/v_2+t-i\epsilon)])^{-\kb},\quad T_0=\frac{v_2}{2\pi a},
\end{align}
\end{widetext}
 with $\alpha=2C$ and $\beta=2\lambda^2S$ defined in Eq. (\ref{eq:ab}), the LL parameters $K_{1/2}$ are defined in Eq. (\ref{eq:luttinger par}). The Fourier transform  $C_{ss}^u(\Delta k,\omega)$ is evaluated using the approximation $\ka\sim0$, %$s_{\pm}\equiv\ x\mp v_2t$ to get:
  %\begin{equation}
 %C_{ss}(\Delta k,\omega)\cong\int ds_+\int ds_-\frac{e^{i(\Delta kx-\omega t)}}{sinh^{\kb}(\frac{\pi T}{v_2}s_+)sinh^{\kb}(\frac{\pi T}{v_2}s_-)}
 %\end{equation}
which simplifies the correlation function into an expression that can be calculated straightforward by the integral \cite{integral}:
\begin{equation}
\int_{0}^{\infty}d\xi \sinh^{-\eta/2}(\pi T\xi)e^{iz\xi}=\frac{2^{\eta-1}}{\pi T}B(\eta/4-\frac{iz}{2\pi T},1-\eta/2)\;
\end{equation}
where $B(x,y)=\frac{\Gamma(x)\Gamma(y)}{\Gamma(x+y)}$ is the Beta function. This yields
\begin{widetext}
\begin{align}
&C_{ss}^u(\Delta k,\omega)=4g_u^2\sin(\pi \kb)\left(\frac{T}{T_0}\right)^{2\kb+2\ka-2}B[\kb/2-\frac{i(\omega/v_2-\Delta k)}{4\pi T},1-K_2\beta^2]B[K_2\beta^2/2-\frac{i(\omega/v_2+\Delta k)}{4\pi T},1-\kb]\; ,
\end{align}
\end{widetext}
and consequently the matrix element
\begin{widetext}
\begin{align}
&M_{ss}^u=-i\partial_{\w}C_{ss}^u|_{\omega=0}=\frac{g_u^2}{2\pi^2}\sin(\pi\kb)\times\nonumber\\
&\left(\frac{T}{T_0}\right)^{2\kb+2\ka-3}B[\kb/2-i\delta,1-\kb]B[\kb/2+i\delta,1-\kb]\Re[\Psi(1-\kb/2-i\delta)-\Psi(\kb/2+i\delta)],\\
&\delta\equiv\frac{v_2\Delta k}{4\pi T}\; .\nonumber
%&g\equiv\frac{J_{\parallel}^{(0)}}{4\pi^2a}\nonumber
\end{align}
\end{widetext}
After deriving the expression for $M_{ss}^u$, we wish to show that $M_{s2}^u$ is proportional to $M_{ss}^u$, and the rest of the elements are found from Eq. (\ref{eq:cons}). Again, using commutator identities we have,
\begin{equation}
C_{s2}^u=-iC_1(x,t)\partial_xC_2(x,t),
\end{equation}
then
\begin{equation}
M_{s2}^u=\int tdt\int dx \Im[C_{s2}^u(x,t)]e^{i\Delta kx}.
\end{equation}
Using $\partial_xe^{2i\beta\phi_2(x,t)}=2i\beta\partial_x\phi_2(x,t)e^{2i\beta\phi_2(x,t)}$ and integrating by parts we have
\begin{widetext}
\begin{equation}
\int dx\partial_xC_2(x,t)e^{i\Delta kx}=-\int dxC_2(x,t)\partial_x(e^{i\Delta kx})=-i\Delta k\int dxC_2e^{i\Delta kx}
\end{equation}
\end{widetext}
which gives the simple relation
\begin{equation}\label{eq:Ms2}
M_{s2}^u=-\frac{1}{4i}(-i\Delta k)M_{ss}^u\Rightarrow M_{s2}^u=\frac{\Delta k}{4}M_{ss}^u\;.
\end{equation}
In a similar way
\begin{equation}\label{eq:M22}
M_{22}^u=(\frac{\Delta k}{4})^2M_{ss}^u\;.
\end{equation}
By substituting Eqs. (\ref{eq:Ms2}), and (\ref{eq:M22}) into Eq. (\ref{eq:cons}) we see that all the elements with $p\text{ or }q=1$ vanish, and the derivation of the umklapp memory matrix is complete.

%checked 18/03 17:09
\section{Disorder Memory Matrix}
\label{sec:app2}
Now we turn to the calculation of the disorder part of the memory matrix. Note that all the non diagonal elements are zero. Since we are interested only in the leading temperature and field dependencies of the memory matrix, the results of the integrals in this section  will be important only to get the powers of $T$ in each matrix element.
%The derivatives of the beta functions with respect to $\omega$ are important because they give a factor proportional to $T^{-1}$.
The force operators is derived from $H_d$ [Eq. (\ref{eq:disorder})], using
\begin{equation}
F_d^u=i[J_p,{H}_d]=i\int dx\zeta(x)[J_p,\cos(2\phi(x))]\; .
\end{equation}
% where the integration is eliminated due to the delta-function in the canonical commutators.
Eq. (\ref{eq:expcorr}) is again useful:  after disorder averaging, and using the identity $\overline{\zeta(x)\zeta(0)}=D\delta(x)$
[Eq. (\ref{eq:zetazeta})], we get
%(for $\epsilon=0$):
\begin{widetext}
\begin{align}
C_{ss}^d(\omega)=\frac{D}{\pi^2}\left(\frac{T}{2T_0}\right)^{\kb/2+\ka/2}\times\nonumber\\
\int dte^{i\omega t}|\sinh(\pi Tt)\sinh(-\pi Tt)|^{-K_2\beta^2/2-\ka/2}=\frac{D}{2\pi^3}\left(\frac{T}{T_0}\right)^{K_2\beta^2/2+\ka/2 -1}B[K_2\beta^2/4-\frac{i\omega}{2\pi T},1-\kb/2]\;.
\end{align}
\end{widetext}
\begin{align}
&M_{ss}^d=-i\frac{\partial C_{ss}^d}{\partial\omega}|_{\omega=0}=D_{ss} T^{K_2\beta^2/2+\ka/2 -2},\nonumber\\
&D_{ss}\cong\frac{D}{\pi v_2}\;.
\end{align}
After some algebra,
\begin{widetext}
\begin{align}
&C_{11}^d(\omega)=D\left(\frac{T}{2T_0}\right)^{K_2\beta^2/2+\ka/2 }\int dx\delta(x)\int dte^{i\omega t}\times\nonumber\\
&\partial_x^2[(\sinh[\pi T(x/v_1-t+i\epsilon)]\sinh[\pi T(x/v_1+t-i\epsilon)])^{-\ka/4}](\sinh[\pi T(x/v_2-t+i\epsilon)]\sinh[\pi T(x/v_2+t-i\epsilon)])^{-\kb/4},%\cong\\
 %\frac{D\pi^2}{2v_1^2}K_1\alpha^2T^{K_2\beta^2/2+K_1\alpha^2/2 +2}\int dt e^{i\omega t}sinh^{-K_2\beta^2/2-K_1\alpha^2/2 -2}
\end{align}
\end{widetext}
\begin{widetext}
\begin{align}
&\Rightarrow C_{11}^d(\omega)\cong \frac{D\ka}{2} \left(\frac{T}{T_0}\right)^{\kb/2+\ka/2+1}B[K_2\beta^2/4+\ka/4+1-\frac{i\omega}{4\pi T},-1-\kb/2-\ka/2]\;.\nonumber
\end{align}
\end{widetext}
\begin{align}
&M_{11}^d=-i\frac{\partial C_{11}^d}{\partial\omega}|_{\omega=0}\cong D_{11} \left(\frac{T}{T_0}\right)^{\kb/2+\ka/2},\nonumber\\
&D_{11}\cong \frac{D\ka }{2v_1}.
\end{align}
The result for $M_{22}^d$ is pretty much the same
\begin{align}
&M_{22}^d=-i\frac{\partial C_{22}^d}{\partial\omega}|_{\omega=0}\cong  D_{22}\left(\frac{T}{T_0}\right)^{K_2\beta^2/2+\ka/2}\nonumber\\
&D_{22}\cong \frac{ D\kb}{2v_2}
\end{align}

\section{Phonon-Spinon Memory Matrix}

In this appendix we detail the calculation of the correlation function appearing in the memory matrix elements responsible for phonon-spinon scattering.
\begin{widetext}
\begin{align}
&C_{p-s}(\w)=\int dt e^{i\w t}\langle F_{p-s}(t);F_{p-s}(0)\rangle=\nonumber\\
&v^4g_{p-s}^2\int dt\sum_{kk'pp'G_n;k_1k_1'p_1p_1'G_{n1}}e^{i(\w+\Delta\w)t}\delta(k-k'+p-p'-G_n)\delta(k_1-k_1'+p_1-p_1'-G_{n1}) pp'(p'-p)p_1p_1'(p_1'-p_1)\times\nonumber\\
&[\langle c_{k'}^{\dag}c_kc_{k_1'}^{\dag}c_{k_1}\rangle\langle b_{p'}^{\dag}b_pb_{p_1'}^{\dag}b_{p_1}\rangle-\langle c_{k_1'}^{\dag}c_{k_1}c_{k'}^{\dag}c_k\rangle \langle b_{p_1'}^{\dag}b_{p_1}b_{p'}^{\dag}b_p\rangle].\nonumber\\
&\Delta\w\equiv\epsilon_k-\epsilon_{k-q}+\w_p-\w_{p+q}.
\end{align}
\end{widetext}
Using Wick's theorem, one obtains
\begin{equation}
\langle\hat{c}_{k'}^{\dag}\hat{c}_k\hat{c}_{k_1'}^{\dag}\hat{c}_{k_1}\rangle=f_kf_{k_1}\delta_{k'k}\delta_{k_1'k_1}-f_kf_{k_1}\delta_{k'k_1}\delta_{k_1'k}+f_{k_1}\delta_{k'k_1}\delta_{k_1'k},
\label{eq:wick}
\end{equation}
and similarly for the other expectation values. We thus get
\begin{align}
&C_{p-s}(\w)=\sum_{kpq} W_{pq}\delta(\w+\Delta\w)n_{p+q}f_{k-q}(1+n_p)(1-f_k),\nonumber\\
&W_{pq}\sim-2v^2g_{p-s}^2|p(p+q)|q^2,\quad q\equiv p'-p.
\end{align}
where $f_k=(e^{(\epsilon_k-\mu)/T}+1)^{-1},\; n_p=(e^{\w_p/T}-1)^{-1}$, are Fermi and Bose  distributions respectively.

%checked 18/03 17:45
%\addcontentsline{toc}{section}{References}

\end{document}